\documentclass[amsmath,amssymb,aps,prb,10pt,superscriptaddress,english,twocolumn,longbibliography]{revtex4-2}
\usepackage{amsmath,amssymb}
\usepackage{graphicx,epsfig}
\usepackage{dcolumn}
\usepackage{color,epstopdf}
\usepackage{tikz}
\usepackage{float}
\usepackage{bm}
\usepackage[normalem]{ulem}
\usepackage[colorlinks=true,bookmarks=false,citecolor=blue,linkcolor=blue]{hyperref}
\usepackage{multirow}
\usepackage{braket}
\usepackage{xcolor}
\usepackage[normalem]{ulem}
\definecolor{Dandelion}{rgb}{0.94, 0.7, 0.19}
\definecolor{dkgreen}{rgb}{0,0.6,0}

\newcommand{\parallelH}{\,\begin{tikzpicture}[scale=0.35,baseline={([yshift=-0.6ex]current bounding box)}]
    \draw (0,0) rectangle (1,1);
    \draw[fill=Dandelion] (0.5,1) ellipse (0.56 and 0.12);
    \draw[fill=Dandelion] (0.5,0) ellipse (0.56 and 0.12);
\end{tikzpicture}\,}
\newcommand{\parallelV}{\, \begin{tikzpicture}[scale=0.35,baseline={([yshift=-.6ex]current bounding box)}]
    \draw (0,0) rectangle (1,1);
    \draw[fill=Dandelion] (0,0.5) ellipse (0.13 and 0.56);
    \draw[fill=Dandelion] (1,0.5) ellipse (0.13 and 0.56);
\end{tikzpicture} \,}
\newcommand{\PlaqOIV}{\, \begin{tikzpicture}[scale=0.35,baseline={([yshift=-.6ex]current bounding box)}]\draw (0,0) rectangle (1,1);
\draw[fill=Dandelion] (1,0.5) ellipse (0.13 and 0.56);
\end{tikzpicture} \,}
\newcommand{\PlaqIOV}{\, \begin{tikzpicture}[scale=0.35,baseline={([yshift=-.6ex]current bounding box)}]\draw (0,0) rectangle (1,1);
\draw[fill=Dandelion] (0,0.5) ellipse (0.13 and 0.56);
\end{tikzpicture} \,}
\newcommand{\emptyPlaq}{\, \begin{tikzpicture}[scale=0.35,baseline={([yshift=-.6ex]current bounding box)}]
    \draw (0,0) rectangle (1,1);
\end{tikzpicture} \,}
\newcommand{\PlaqOIH}{\,\begin{tikzpicture}[scale=0.35,baseline={([yshift=-0.6ex]current bounding box)}]
    \draw (0,0) rectangle (1,1);
    \draw[fill=Dandelion] (0.5,0) ellipse (0.56 and 0.12);
\end{tikzpicture}\,}
\newcommand{\PlaqIOH}{\,\begin{tikzpicture}[scale=0.35,baseline={([yshift=-0.6ex]current bounding box)}]
    \draw (0,0) rectangle (1,1);
    \draw[fill=Dandelion] (0.5,1) ellipse (0.56 and 0.12);
\end{tikzpicture}\,}

\newcommand{\ohdimer}{\begin{tikzpicture}[scale=0.20,baseline={([yshift=-0.7ex]current bounding box.north)}]
    \filldraw (0,1) circle (7pt);
    \draw (0,1) -- (1.5,1);
\end{tikzpicture} }

\newcommand{\hdimero}{
\begin{tikzpicture}[scale=0.20,baseline={([yshift=-0.7ex]current bounding box.north)}]
    \filldraw (1.5,1) circle (7pt);
    \draw (0,1) -- (1.5,1);
\end{tikzpicture} }
\newcommand{\ovdimer}
{\,
\begin{tikzpicture}[scale=0.2,baseline={([yshift=-0.1ex]current bounding box.south)}]
    \filldraw (0.5,0.5) circle (7pt);
    \draw (0.5,0.5) -- (0.5,1.5);
\end{tikzpicture}
}

\newcommand{\vdimero}
{\,
\begin{tikzpicture}[scale=0.2,baseline={([yshift=-0.1ex]current bounding box.south)}]
    \filldraw (0.5,1.5) circle (7pt);
    \draw (0.5,0.5) -- (0.5,1.5);
\end{tikzpicture}
}
\begin{document}

\title{Re-entrance effect in the high-temperature critical phase of the quantum dimer model on the square lattice}

\author{Bhupen Dabholkar}
\email{dabholkar@irsamc.ups-tlse.fr}
\affiliation{Laboratoire de Physique Th\'eorique, Universit\'e de Toulouse, CNRS, UPS, France}

\author{G.J. Sreejith}
\email{sreejith@acads.iiserpune.ac.in}
\affiliation{IISER Pune, Dr Homi Bhabha Road, Pune 411008, India}

\author{Fabien Alet}
\email{fabien.alet@cnrs.fr}
\affiliation{Laboratoire de Physique Th\'eorique, Universit\'e de Toulouse, CNRS, UPS, France}

\date{\today}

\begin{abstract}
We present a quantum Monte Carlo investigation of the finite-temperature phase diagram of the quantum dimer model on the square lattice. We use the sweeping cluster algorithm, which allows exact implementation of the dimer constraint, supplemented with an equal-time directed loop move that allows sampling the winding sectors. We find a high-temperature critical phase with power-law correlations that extend down to the Rokshar-Kivelson point, in the vicinity of which a re-entrance effect in the lines of constant exponent is found. For small values of the kinetic energy strength, we find finite-temperature transitions to ordered states (columnar and staggered) which match those of interacting classical dimer models.
\end{abstract}

\maketitle

\section{Introduction}
\label{sec:intro}
In many physical systems, widely differing energy scales can cause the system to explore only a subset of all the allowed configurations. 
The essential low energy physics of the true many-body Hamiltonian is then well described by an effective and sometimes more abstract constrained model. 
This is the case, for instance, in frustrated magnets, where the system can only be in a sub-manifold of degenerate constrained configurations.  
The quantum dimer model (QDM), originally introduced in the context of high-temperature superconductivity~\cite{Rokhsar88}, is the prototypical example of a constrained quantum model. 
It is relevant as a low-energy effective description of frustrated antiferromagnets~\cite{Moessner2011}, in atomic physics of Rydberg arrays~\cite{Semeghini2021,Verresen2021,Yan2022a,Rhine} and lattice gauge theories~\cite{Moessner2001c}. 

The QDM on the two dimensional square lattice describes the physics of hardcore fully packed dimers that live on the edges between nearest neighbor sites of the lattice.

The zero-temperature phase diagram of the QDM, showcasing the competition between a kinetic energy (of strength denoted by $t$ hereafter, see Eq.\ref{eq:HQDM} below for the concrete description of the QDM) and a potential energy (of strength denoted by $V$) has been studied in a variety of different lattice geometries~\cite{Sachdev89,Leung96,Moessner2001,Moessner2001b,Misguich2002,Misguich2003,Ralko2005,Syljuasen2005,Vernay2006,Ralko2008,Sikora2009,Sikora2011,Banerjee2014,Buerschaper2014,Schlittler2015,Qi2015,Banerjee2016,Oakes18,Yan2021,Yan2021b}. 
The square lattice case, in particular, has been the subject of many studies. 
For $V/t>1$, it is agreed that the ground-state is highly degenerate and consists of single classical dimer configurations with no flippable plaquettes (see a discussion in Ref.~\onlinecite{Mambrini2015} and Sec.~\ref{sec:obs_stag} on this degeneracy). On the square lattice, flippable plaquettes denote plaquettes where two parallel dimers are present on opposite edges (see Eq.~\ref{eq:HQDM}). The ground-state at $V/t=1$, called the Rokhsar-Kivelson (RK) point~\cite{Rokhsar88}, takes the simple form of a uniform superposition of all dimer configurations. 
This allows us to make use of results from classical statistical-mechanics~\cite{Fisher63}, showing that correlations between dimers are power-law decaying at the critical RK point. At large enough negative $V/t$, a columnar ordered ground-state (breaking discrete lattice symmetries) is expected. 
The nature of the intermediate regime between the Rokhsar Kivelson point and the large negative $V/t$ regions is still debated; different phases containing columnar, mixed and/or plaquette phases have been proposed in this regime (see Ref.~\onlinecite{Yan2021} for a recent review). The existence of a pseudo-rotor spectrum (associated to an emergent U(1) symmetry) close to the RK point~\cite{Banerjee2014,Oakes18} complicates the analysis of simulations for $V/t \lesssim 1$.

The present work is concerned with the finite-temperature $(T>0)$ phase diagram of the QDM, for which there is no previous microscopic study to the best of our knowledge. A lot of intuition can be drawn from the finite-temperature phase diagram of a {\it classical} interacting dimer model (CIDM), which has been analyzed extensively in the past using numerical simulations and field theory~\cite{Alet2005,Alet2006,Papanikolaou2007,AletHDR,Wilkins2020}. 
The CIDM is equivalent to setting the kinetic energy strength $t$ of the QDM to $0$, and analyzing the behavior of the QDM as a function of the temperature $T/|V|$. The  phases realized in the CIDM are as follows.
For both signs of $V$ (attractive or repulsive interactions), there is a high-temperature critical phase (above a transition temperature $T_c(V)$) where the dimer-dimer correlations display a power-law behavior with a temperature dependent exponent. 

\begin{figure*}[t]

\includegraphics[width=0.99 \textwidth]{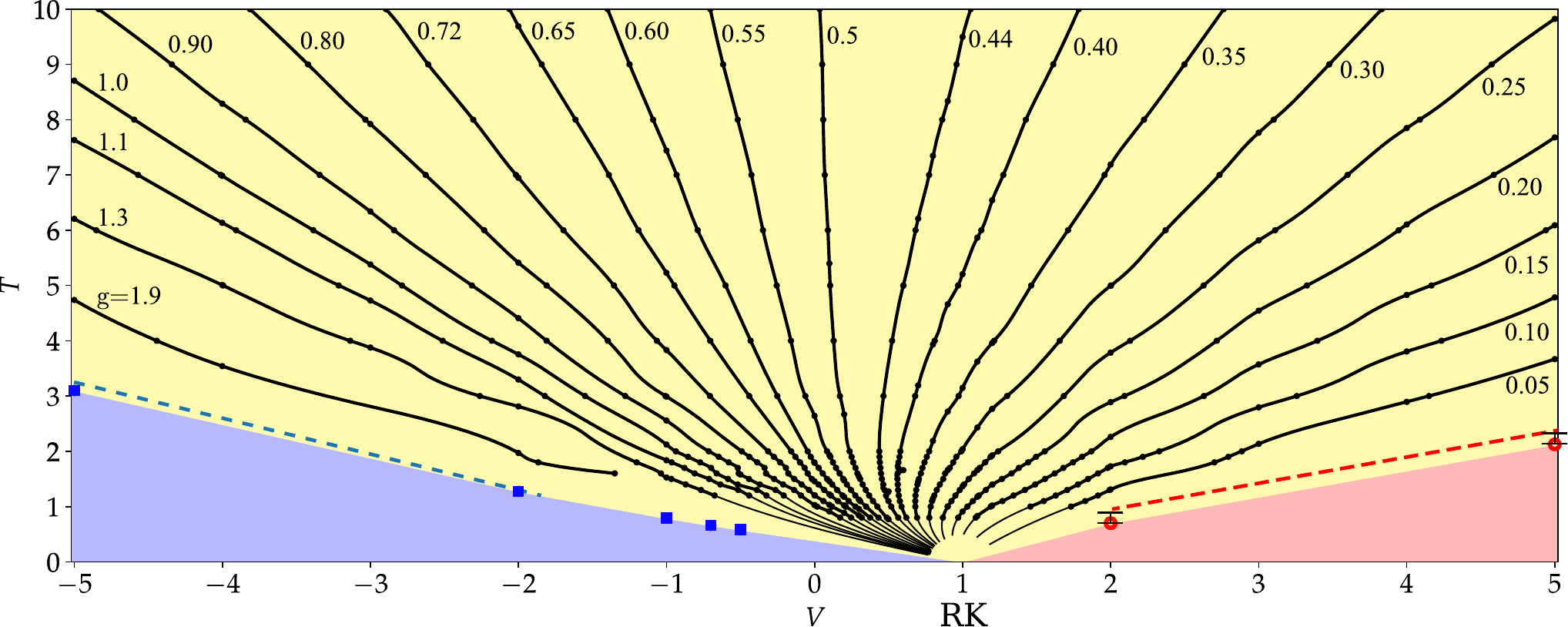}

\caption{\label{fig:pd} Finite-temperature phase diagram of the quantum dimer model on the square lattice, in the ($V/t$, $T/t$) plane ($t=1$ is assumed here). The yellow, blue and red shades indicate the critical (disordered) phase, the columnar phase and the staggered phase respectively. The black solid lines show constant $g$ lines. The black dots were obtained by linear interpolation of $g$  values calculated using Eq. \ref{eq:winding} from numerically obtained $\langle W^2\rangle$ values on a large grid of $(V,T)$ values (see Sec.~\ref{sec:results2}). The thick solid lines joining them are quadratic interpolations of the dots with same $g$. Thin solid lines are extrapolations towards the RK point $V/t=1$; these regions cover regions inaccessible for the QMC simulations. To map out the entire high-temperature phase, we used simulations for systems of linear size $L=24$ (we checked on selected points that increasing system size up to $L=64$ did not change the phase diagram significantly). The blue (square) and red (empty circle) symbols denote the finite-temperature phase transitions to a columnar phase and staggered phase. These are determined using, respectively, the Binder cumulants of the dimer symmetry breaking order parameter $B_D$ (Sec.~\ref{sec:results1}) and the absolute winding $B_{|W|}$ (Sec.~\ref{sec:results3}). The dotted lines represent the classical asymptotic behavior obtained for the CIDM where $T_c=0.65 |V|$ (for $V<0$, attractive CIDM~\cite{Alet2006}) and $T_c=0.475 V$ (for $V>0$, repulsive CIDM~\cite{Wilkins2020}). The transition to columnar (respectively staggered) order  are expected to occur at $g_c=4$ (respectively $g=0$).}
\end{figure*}

For the most well-studied case of attractive interactions ($V<0$) the low-temperature phase is columnar (with a four-fold degenerate ground-state) and the transition is of Kosterlitz-Thouless type~\cite{Alet2005,Alet2006,Papanikolaou2007}. For repulsive interactions ($V>0$) the low-temperature phase is of ``staggered'' type with, however, a large degeneracy as mentioned above. The nature of the transition is slightly less well established here: analytical arguments suggest an infinite-order phase transition showing both first-order and continuous transition characteristics while numerical simulations only observe continuous aspects~\cite{AletHDR,Wilkins2020}. All these features can be understood at a coarse-grained level from an effective height field theory described by the 2D sine Gordon action (derived heuristically for the repulsive case~\cite{Alet2005,Alet2006})
\begin{equation}
    S = \int d^{2}r[ \pi g (\nabla h)^{2} + \sum_{p=4,8 \dots} {\cal V}_p \cos\left(2\pi p h\right) ]
    \label{eq:Saction}
\end{equation}
where $h(r)$ is the compact height field with period $1$. The first elastic term accounts for the height field, and is associated with the constrained fluctuations of the dimer configurations, while the cosine terms tend to favor dimer ordering, which for the most relevant operator at $p=4$ describes the columnar ground states (when ${\cal V}_4>0$). The ``Coulomb gas'' constant $g$ varies with the temperature $T/|V|$ of the microscopic model and determines, in part, the critical dimer-dimer correlations with a power-law contribution decaying with an exponent $1/g$. For $V<0$, as the temperature $T/|V|$ is lowered, $g$ increases from $1/2$ for $T/|V|\rightarrow \infty$ to $g_c(T_c)=4$ on approaching the critical temperature $T_c$ for the transition into the low temperature columnar ordered phase; at $g=4$, the cosine term becomes relevant and drives the system into the columnar ordered state.
For $V>0$, $g$ is expected to decrease with temperature $T/|V|$ until $g\rightarrow 0$, where the subleading terms (not included in Eq.~\ref{eq:Saction}) become important and could describe the transition to the non-flippable states~\cite{Otsuka2009,AletHDR,Wilkins2020}.  We note that Eq.~\ref{eq:Saction} can alternatively be obtained from a quantum treatment of a QDM field theory at finite temperature (see Ref.~\onlinecite{Szabo2019,Nogueira09} for different approaches and for explicit expressions for the subleading terms), and was justified for the CIDM using rigorous arguments in Ref.~\onlinecite{Giuliani17}.

Based on the equivalence between quantum models at finite temperature and classical models, we expect that Eq.~\ref{eq:Saction} describes the high-temperature phase of the QDM, including the transitions to the low-temperature ordered phases. In this work, we numerically study, using quantum Monte Carlo (QMC) techniques, the finite-temperature phase diagram of the QDM on the square lattice and verify this expectation, focusing in particular on the high-temperature critical phase parametrized by a $g$ that depends non-trivially on both the temperature and the potential $V$.
Reference~\onlinecite{Alet2006} conjectured a specific form for the phase diagram (see Fig.~32 in Ref.~\onlinecite{Alet2006}) with monotonous behavior of the ``iso-$g$'' lines. We show in our work that the actual constant $g(T,V)$ lines exhibit a re-entrant behavior as a function of $T$ below the RK point. Additionally, we investigate the phase transitions (starting from the high-temperature critical phase) deep in the columnar ($V/t <  1$) and staggered-like ($V/t > 1$) regimes, and conclude that they are similar to the ones observed in the CDIM obtained in the large potential limits ($|V/t| \rightarrow \infty$). These results are summarized in the phase diagram of Fig.~\ref{fig:pd}.

Our manuscript is organized as follows. 
Section~\ref{sec:sec2} contains the definition of the model (Sec.~\ref{sec:model}), the simulation method (Sec.~\ref{sec:method}) and relevant observables (Sec.~\ref{sec:obs}) to describe the phase diagram. 
Section~\ref{sec:results} presents our results in the large attractive potential region where we observe a transition to columnar order (Sec.~\ref{sec:results1}), in the intermediate potential region focusing on the properties of the high-temperature phase (Sec.~\ref{sec:results2}), and finally in the large repulsive potential region which hosts a transition into a staggered phase (Sec.~\ref{sec:results3}). Conclusions and perspectives are presented in Sec.~\ref{sec:conc}. The Appendix~\ref{sec:appendix-MC} contains the details of the QMC sweeping cluster algorithm used to simulate the QDM, and in particular the equal-time directed loop move needed to perform ergodic simulations at finite temperature. The Appendix~\ref{sec:app-highT} contains an argument based on high-temperature expansions of the QDM and CIDM which provides an excellent account of  the high-temperature part of the phase diagram Fig.~\ref{fig:pd}.

\section{Model, observables and simulation details}
\label{sec:sec2}

\subsection{Model}
\label{sec:model}
The quantum dimer model on the square lattice is described by the Hamiltonian:
\begin{widetext}
\begin{equation}
H_{\rm QDM}   =  \sum_{\square} -t  \left( |
\begin{tikzpicture}[scale=0.35,baseline={([yshift=-0.6ex]current bounding box)}]
    \draw (0,0) rectangle (1,1);
    \draw[fill=Dandelion] (0.5,1) ellipse (0.56 and 0.12);
    \draw[fill=Dandelion] (0.5,0) ellipse (0.56 and 0.12);
\end{tikzpicture}
\rangle \langle
\begin{tikzpicture}[scale=0.35,baseline={([yshift=-.6ex]current bounding box)}]
    \draw (0,0) rectangle (1,1);
    \draw[fill=Dandelion] (0,0.5) ellipse (0.13 and 0.56);
    \draw[fill=Dandelion] (1,0.5) ellipse (0.13 and 0.56);
\end{tikzpicture} | + |
\begin{tikzpicture}[scale=0.35,baseline={([yshift=-.6ex]current bounding box)}]
    \draw (0,0) rectangle (1,1);
    \draw[fill=Dandelion] (0,0.5) ellipse (0.13 and 0.56);
    \draw[fill=Dandelion] (1,0.5) ellipse (0.13 and 0.56);
\end{tikzpicture}
\rangle \langle
\begin{tikzpicture}[scale=0.35,baseline={([yshift=-0.6ex]current bounding box)}]
    \draw (0,0) rectangle (1,1);
    \draw[fill=Dandelion] (0.5,1) ellipse (0.56 and 0.12);
    \draw[fill=Dandelion] (0.5,0) ellipse (0.56 and 0.12);
\end{tikzpicture} |
\right) + V   \left( |
\begin{tikzpicture}[scale=0.35,baseline={([yshift=-0.6ex]current bounding box)}]
    \draw (0,0) rectangle (1,1);
    \draw[fill=Dandelion] (0.5,1) ellipse (0.56 and 0.12);
    \draw[fill=Dandelion] (0.5,0) ellipse (0.56 and 0.12);
\end{tikzpicture}
\rangle \langle
\begin{tikzpicture}[scale=0.35,baseline={([yshift=-0.6ex]current bounding box)}]
    \draw (0,0) rectangle (1,1);
    \draw[fill=Dandelion] (0.5,1) ellipse (0.56 and 0.12);
    \draw[fill=Dandelion] (0.5,0) ellipse (0.56 and 0.12);
\end{tikzpicture}
 | + |
\begin{tikzpicture}[scale=0.35,baseline={([yshift=-.6ex]current bounding box)}]
    \draw (0,0) rectangle (1,1);
    \draw[fill=Dandelion] (0,0.5) ellipse (0.13 and 0.56);
    \draw[fill=Dandelion] (1,0.5) ellipse (0.13 and 0.56);
\end{tikzpicture}
\rangle \langle
\begin{tikzpicture}[scale=0.35,baseline={([yshift=-.6ex]current bounding box)}]
    \draw (0,0) rectangle (1,1);
    \draw[fill=Dandelion] (0,0.5) ellipse (0.13 and 0.56);
    \draw[fill=Dandelion] (1,0.5) ellipse (0.13 and 0.56);
\end{tikzpicture} |
\right)
\label{eq:HQDM}
\end{equation}
\end{widetext}
where $t$ denotes the strength of the kinetic energy term that flips plaquettes with parallel dimers, and $V$ the strength of the potential term which counts the flippable plaquettes. We will often set $t=1$ such that all energy scales (temperature $T$, potential strength $V$) are considered in units of $t$, if not specified.
We will also often refer to the classical interacting dimer model obtained by setting $t=0$. For the CIDM, we always consider $|V|=1$ and only the sign of $V$ matters, with $V=-1$ ($V=1$) corresponding to attractive (repulsive) interactions between dimers. We consider square lattices of linear size $L$, hosting $L^2$ sites, $2L^2$ bonds and $L^2/2$ dimers and use periodic boundary conditions.

\subsection{Simulations}
\label{sec:method}

Our QMC simulations are based on the recently developed sweeping cluster algorithm \cite{Yan2019} which is able to efficiently sample systems with geometric constraints such as the QDM (see Refs.~\onlinecite{Yan2022a,Yan2022b,Yan2022c,Ran2022} for application to other constrained models). The main asset of the algorithm is the ability to sample the path integral by satisfying the hardcore geometric constraint at each (infinitesimal) imaginary time slice, corresponding thus to a perfect representation of the QDM. However, one drawback of the sweeping cluster algorithm as originally introduced is that it is not fully ergodic: it only samples one winding sector (see definition in Sec.~\ref{sec:obs2}), leading to unreliable results when studying finite temperature properties of the model. As seen later, the winding fluctuations are crucial to understand the physics at high temperature. A simple improvement which we use in this work is to add a global equal-time directed loop move, which performs a loop update on a time slice -- allowing to change the winding sector of the dimer configuration. This update is performed under the constraint that all time slices can accept such a loop move, and a unit acceptance probability is ensured for this update. Details on how to build efficiently this loop move are provided in Appendix~\ref{sec:appendix-MC}. As we go to lower temperatures, it becomes harder to make such a global move and the algorithm starts freezing in a particular winding sector. We can nevertheless reach equilibration of sufficiently large systems at intermediate to low temperatures. We note that a different solution has been proposed in Ref.~\onlinecite{Yan2022}, which incorporates such a loop move {\it inside} the construction of the sweeping cluster. It is however not clear that this latter move allows to reach an ergodic sampling (with a reasonable acceptance rate) on large systems at intermediate or high temperatures, which is the main focus of the current work.

\subsection{Observables}
\label{sec:obs}

\subsubsection{Observables to monitor the phase transition to crystalline columnar-like ordering}
\label{sec:obs1}
As discussed in the introduction, the ground-state of the QDM on the square lattice for $V/t<1$ has been argued to be of columnar, plaquette or mixed type -- breaking different symmetries. The standard observable to distinguish these phases is the complex order parameter $\Psi_{\rm col}$ introduced in Ref.~\onlinecite{Sachdev89}. It is defined locally at lattice site ${\bf r}$ as:
$\Psi_{\rm col} ({\bf r}) = (-1)^x\left[
d^{\,\ohdimer}({\bf r}) - 
d^{\hdimero}({\bf r}) \right] + \dot{\imath}\,(-1)^y\left[
d^{\ovdimer}({\bf r}) - 
d^{\vdimero}({\bf r}) \right]
$
where ${\bf r}\equiv (x,y)$ the dot $\begin{tikzpicture}[scale=0.2,,baseline={([xshift=+6ex,yshift=-0.1ex]current bounding box.south)}]\filldraw (0.0,0.0) circle (7pt);\end{tikzpicture}$ and the line 
$\begin{tikzpicture}[scale=0.2,,baseline={([xshift=+6ex,yshift=-0.5ex]current bounding box.south)}]\draw (0.0,0.0) -- (1.0,0.0);\end{tikzpicture}$
denotes the relative positions of the lattice site ${\bf r}$ and an adjoining edge. The  dimer operator $d^{\,\ohdimer}({\bf r})$ is $1$ if there is a horizontal dimer on the right horizontal bond touching lattice site ${\bf r}$, and $0$ otherwise (other operators are defined similarly). The total order parameter is averaged over all sites belonging to one sublattice $A$ of the square lattice  sites $\Psi_{\rm col}=\frac{2}{L^2}\sum_{{\bf r} \in A} \Psi_{\rm col} ({\bf r})$.

The magnitude of the columnar order parameter $|\Psi_{\rm col} |$ acquires a non-zero value in columnar, plaquette and mixed phases, allowing us to distinguish these from the disordered phase. On the other hand,  $\theta=\arg(\Psi_{\rm col})$ allows us to distinguish between the these phases (see e.g. discussion in Ref.~\onlinecite{Leung96,Yan2021}). This has been used to argue that the QDM admits a weakly mixed ordered phase at zero-temperature~\cite{Yan2021}, with a very tiny signal visible in the histograms of $\theta$.

To avoid related statistical issues associated with a very weak signal, we also consider the dimer rotation symmetry breaking parameter 
 $D = 2/L^2 \langle \left| N( 
\begin{tikzpicture}[scale=0.4,baseline={([yshift=-0.6ex]current bounding box)}]
    \draw[fill=Dandelion] (0.5,1) ellipse (0.56 and 0.12);
\end{tikzpicture} 
 ) -  N( \,
\begin{tikzpicture}[scale=0.4,baseline={([yshift=-.6ex]current bounding box)}]
    \draw[fill=Dandelion] (0,0.5) ellipse (0.13 and 0.56);
\end{tikzpicture}
 \, ) \right|\rangle $, which measures in every configuration the absolute value of the difference between the number of horizontal versus vertical dimers. $D$ captures the broken $\pi/2$ rotational symmetry in the columnar (or mixed) phase for $V/t < 1$, but vanishes in a plaquette phase. We measure both $\langle D \rangle$ and the associated Binder cumulant $B_D={\langle D^4 \rangle}/{\langle D^2\rangle^2}$. 
 We note, as first discussed in Ref.~\cite{Papanikolaou2007}, that the Binder cumulant should not exhibit a crossing point (with different system sizes) but instead $B_D(T)$ for different $L$'s should coincide below the KT transition temperature, while splitting away from each other above. However strong finite-size effects allowed the use of crossing points of $B_D$ to pinpoint the phase transition in the CIDM~\cite{Alet2005,Alet2006,Papanikolaou2007}. We will also exploit this fact for the study of the finite-temperature phase transition into the low temperature columnar/mixed phase in the QDM.

\subsubsection{Observables to monitor the critical high-T phase}
\label{sec:obs2}

The high-temperature phase is characterized by winding number fluctuations as well as power-law dimer-dimer and monomer-monomer correlations. Let us define these objects and their Monte Carlo measurements.

{\it Winding number fluctuations --- }
On a square lattice, the winding numbers $(W_{x}, W_{y})$ are defined as follows: consider a given dimer configuration and a reference line (a loop, when using periodic boundary conditions) in, say the $y$ direction. Due to the bipartiteness of the lattice, we can divide the bonds which cross this reference line into even and odd bonds. Then the winding number, say $W_{x}$, is defined as $N_{o} - N_{e}$, where $N_{o}$ ($N_{e}$) for a given configuration is the number of dimers on the odd (even) bonds crossing the $y$ reference line. The individual winding numbers $W_x,W_y$ can take values between $-L/2$ and $L/2$, and the dimer configurations are such that the allowed pairs $(W_x,W_y)$ are located within the rhombus defined by $|W_x|+|W_y|\leq L/2$ (see Ref.~\onlinecite{Mambrini2015,Wilkins2020} for illustration of this rhombus).

The winding number fluctuations $\langle W^{2} \rangle = \frac{1}{2}(\langle W_{x}^{2}\rangle + \langle W_{y}^{2}\rangle )$ are particularly useful to characterize the critical phase at finite temperature. Indeed it was shown \cite{Alet2005} that there is a direct relation between $\langle W^{2} \rangle$ and the Coulomb gas constant $g$ of Eq.~\ref{eq:Saction}: 
\begin{equation}\label{eq:winding}
    \langle W^{2} \rangle = \sum_{n \in \mathbb{Z}}n^{2}e^{-g\pi n^{2}} / \sum_{n \in \mathbb{Z}}e^{-g\pi n^{2}}.
\end{equation}

This relation holds as long as the ``rough'' gas phase is stable, that is for $0<g\leq g_c$. It offers a way to measure the Coulomb gas constant easily by simply monitoring the winding fluctuations in the Monte Carlo simulation. The method allows us to map the $g$ as a function of $T$ and $V$ in a large region of the high temperature phase of the QDM. 
Of particular interest are the values at the free-fermion point $g=1/2$ and the expected critical value $g_c=4$ for columnar ordering. $g=1/2$ corresponds to the dimer model with no interactions (that is, the infinite-temperature limit of both classical and quantum dimer models), and $\langle W^{2} \rangle \simeq 0.30343 $ there. On the other hand, the winding fluctuations take an extremely small value $\langle W^{2} \rangle \simeq \mathcal{O}(10^{-6})$
at $g_c=4$. It is not feasible to reach a level of accuracy using Monte Carlo simulations that will be able to resolve this from the value $\langle W^2 \rangle=0$ in the low-temperature columnar (or mixed) phase. Already for $g=2$, we have $\langle W^2 \rangle \sim \mathcal{O}(10^{-3})$. This explains why we cannot effectively map out $g$ values in the regions close to the columnar phase transitions where we expect larger values of $g$ (See Fig \ref{fig:pd}).

{\it Dimer-Dimer correlations --- }
In order to further understand the high-temperature regime, we monitor the behavior of equal-time parallel dimer-dimer correlations. The longitudinal dimer dimer correlation measures the correlation between dimer occupancies in two edges that are both collinear with their separation. It is defined as 
$
G^l(x)= \langle d^{\,\ohdimer}({\bf r}) d^{\,\ohdimer}({\bf r}+(x,0) \rangle^c
.$ The distance vector $(x,0)$ separating the two edges is along the horizontal direction.
The transverse correlator is 
$G^t(x)=\langle d^{\,\ohdimer}({\bf r}) d^{\,\ohdimer}({\bf r}+(0,x) \rangle^c
$
where the distance vector $(0,x)$ is along the vertical direction. 
Assuming rotational and translation symmetry, we can average each correlator over the two perpendicular directions and over all positions $\bf r$. Note that we consider the connected correlator $\langle \dots \rangle^c$, where the square of the average dimer occupation $\langle d^{\,\ohdimer}\rangle^2 = 1/16$
is subtracted. 
These equal-time correlators are easily accessed in the QMC simulations as they correspond to diagonal operators in the computational basis. We average the correlator over all time-slices to improve statistics.

An analysis~\cite{Papanikolaou2007,Sandvik2006} of the operator content of the dimer operator in the continuum limit (Eq.~\ref{eq:Saction}) shows that in the critical phase $0<g<g_c$, the dimer-dimer correlators contain two leading contributions: one scaling with a fixed power-law $1/x^2$ (the so-called dipolar contribution), and one (the vertex contribution) with an exponent related to the Coulomb gas constant $1/x^{1/g}$. 
The long-distance behavior is thus dominated by the contribution with the smallest exponent $\min (2,1/g)$. 
Both contributions may have a prefactor that may depend on $x$, but does not scale with $x$. 
For instance, it was remarked~\cite{Alet2006} that in the attractive CIDM (for which $g\leq 1/2$ and therefore the  contribution $1/x^{1/g}$ displays the slowest decay), the longitudinal correlator has a staggered leading contribution $G^l(x)\propto (-1)^x x^{-1/g}$.

On a lattice of linear size $L$ with periodic boundary conditions, the correlators $G^l(x)$ and $G^t(x)$ are symmetric around $x=L/2$. Anticipating the validity of Eq.~\ref{eq:Saction} and power-law decays, we will often perform the conformal rescaling $x \rightarrow \tilde{x} = \frac{L}{\pi} \sin (\pi x /L) $ and fit correlators in the range $x \in [1,L/5]$.

{\it Monomer-monomer correlators --- } The final correlation function we consider is the (equal-time) monomer-monomer correlator $M({\bf x})=\langle m({\bf r}) m({\bf r}+{\bf x}) \rangle$, where $m({\bf r})$ is $1$ if there is a monomer at lattice site ${\bf r}$ and $0$ otherwise. A monomer is defined as a lattice site with no dimer. While the QDM Hilbert space does not allow monomers, one can define $M({\bf x})$ in an extended space where two test monomers are allowed. At the classical level, $M({\bf x})$ denotes the number of dimer configurations, weighted by their Boltzmann weights, which allow for the two monomers separated by a distance ${\bf x}$ (two monomers are necessarily on opposite sublattices on the square lattice). $M({\bf x})$ can be measured thanks to intermediate configurations in the Monte Carlo loop algorithm~\cite{Alet2006,Sandvik2006,Alet2003} which indeed works in the extended ensemble with two monomers. We adapt this strategy as the equal-time directed loop move described in Appendix~\ref{sec:appendix-MC} is similar and allows to identically estimate the equal-time correlator $M({\bf x})$ for the QDM. 

The monomer-monomer correlator has also been studied in the CIDM, and it was shown~\cite{Alet2005,Alet2006,Papanikolaou2007} that it is expressed in the continuum theory through a magnetic vertex operator, and decays as $1/x^{g}$. We will primarily consider the case ${\bf x}=(0,x)$ (averaging data with the symmetric orientation $(x,0)$) and also use conformal rescaling to account for the symmetry of $M(x)$ around $x=L/2$ and fit for this expected power-law behavior.

\subsubsection{Observables to monitor the staggered phase and phase transition}
\label{sec:obs_stag}

When $V/t > 1$, the ground-state of the QDM has an exactly vanishing energy and is highly degenerate. All configurations that have no flippable plaquette are ground-states. Besides the four staggered configurations for which $|W_x|$ or $|W_y|$ equals $L/2$, all configurations for which $|W_x|+|W_y|=L/2$ are also ground states~\cite{Mambrini2015,Wilkins2020}. We therefore define an order parameter $|W|=|W_x|+|W_y|$ and calculate its Binder cumulant through $B_{|W|}={\langle |W|^4 \rangle}/{\langle |W|^2 \rangle^2}$ as it was shown~\cite{AletHDR,Wilkins2020} to be efficient to detect the transition to the staggered phase in the case of the CIDM with repulsive interactions $V=1$. 

\section{Results}
\label{sec:results}

The results presented in the sections below allow us to draw the phase diagram of the QDM in the $(V/t,T/t)$ plane of Fig.~\ref{fig:pd}.

\subsection{Negative $V/t$ range: transition to a columnar phase}
\label{sec:results1}

\begin{figure}[t]
\includegraphics[width=0.99 \columnwidth]{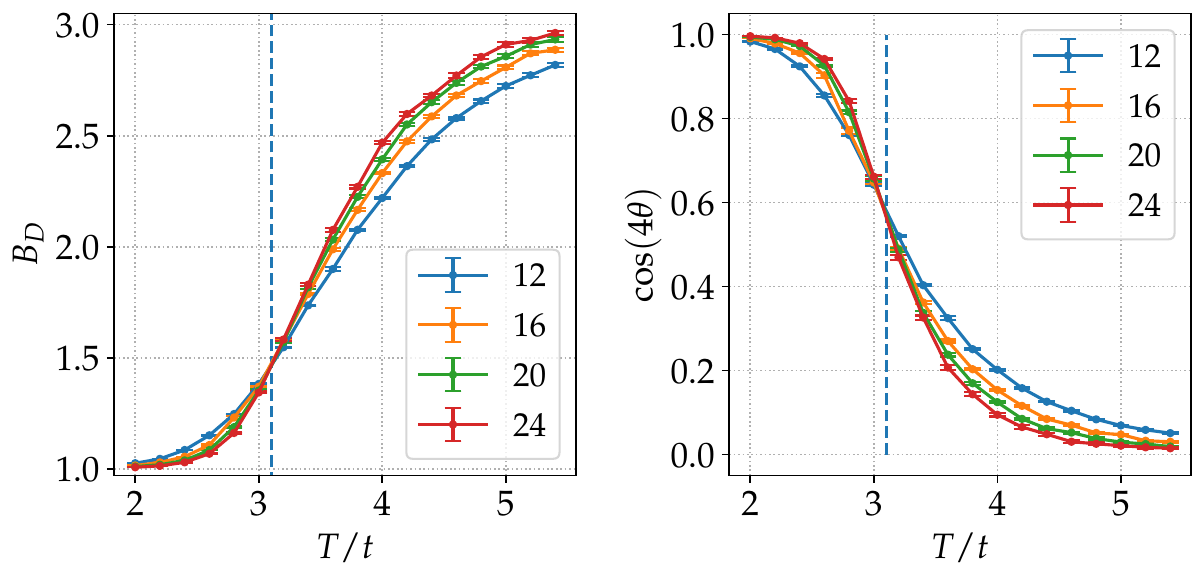}
\caption{\label{fig:columarorder} Finite-temperature columnar ordering transition for $V/t=-5$ and different linear lattice sizes $L$. a) Binder cumulant of $D$ showing a crossing point for different system sizes. b) Similar behavior for $\langle \cos(4 \theta) \rangle $ where $\theta = \arg(\Psi_{\rm col})$ is the phase of the columnar order parameter defined in Sec.~\ref{sec:obs1}. The angle $\theta$ is multiplied by $4$ such that all columnar states characterized by $\theta=0,\pi/2,\pi,3\pi/2$ contribute to produce the same value $\cos( 4 \theta ) = 1$ at low temperature. Crossing points of $B_D$ and $\langle \cos(4 \theta) \rangle $ give a similar estimate $T_c \simeq 3.1 t$ denoted by the dashed line.
}
\end{figure}

We study the transition into the low-temperature ordered phase in the $V/t<0$ regime by monitoring the Binder cumulant $B_D$ (see Sec.~\ref{sec:obs1}). Representative data obtained for $V/t=-5$ are presented in Fig.~\ref{fig:columarorder}a, where we observe a crossing point for different system sizes $L \in [12,24]$ signaling a transition at an estimated $T_c \simeq 3.1 t $ for this value of $V/t$. We obtain similar results for different values of $V/t$, resulting in the blue markers of Fig.~\ref{fig:pd}. In analogy with the classical case, we expect this phase transition to be of Kosterlitz-Thouless type. We note that the value of $B_D$ at the critical point changes with $V/t$ (not shown), in agreement with the fact that this is not a second-order phase transition and that the Binder cumulant crossing is due to subleading finite-size effects~\cite{Papanikolaou2007}. For $V/t\in ]-0.5,1[$, we are not able to reach equilibration of QMC simulations down to the temperature range where the system starts to order, however we expect that there is a similar finite-temperature phase transition for all values $V/t<1$.

An important remark is that the  dimer rotation symmetry breaking order parameter $D$, is non-zero for both columnar and mixed states. In principle, the finite-temperature phase transition could also be to a mixed phase. However, we find that in the range $V/t \leq -0.5$  and down to the lowest temperatures which we could reach while maintaining the winding fluctuations, the phase of the columnar order parameter $\theta=\arg (\Psi_{\rm col})$ (see Sec.\ref{sec:obs1} and References~\onlinecite{Sachdev89,Leung96,Yan2021}) indicates a columnar ordering within error bars. The actual behavior of $\cos(4 \theta)$ for $V/t=-5$ is represented in Fig.~\ref{fig:columarorder}b and also displays a crossing point at the phase transition temperature $T_c \simeq 3.1 t$: at low temperature, all curves tend to reach $\langle \cos( 4 \theta ) \rangle = 1$ , while in the high-temperature phase with no symmetry breaking $\cos(4 \theta)$ averages to $0$.  Also, we expect that the columnar instability (caused by a positive value of ${\cal V}_4$ in the action Eq.~\ref{eq:Saction} leading to a critical value $g_c=4$) should occur first as the $p=4$ operator is the most relevant term. Unfortunately, we cannot conclude from our numerics alone that the value of $g$ at criticality is indeed $g_c=4$, due to the very small values of $\langle W^2 \rangle$ it implies (see Sec.~\ref{sec:obs2}).  Overall, the results we have suggest that all finite-T transitions that we observe in the negative $V/t$ range are to the columnar ordered phase. We cannot make any definitive statement for $V/t \in ] -0.5,1[$, {\it i.e.} closer to the RK point. Also, we cannot rule out occurrence of another phase transition at a lower temperature (lower than the columnar transition temperature) towards a mixed state. This could occur for instance if  ${\cal V}_4$ vanishes. In the absence of equilibrated QMC simulations at low but finite temperature, this can only remain a speculation.

\subsection{Intermediate $V/t$ range: the re-entrant behavior in the high-T phase }
\label{sec:results2}

\begin{figure}[t]

\includegraphics[width=0.99 \columnwidth]{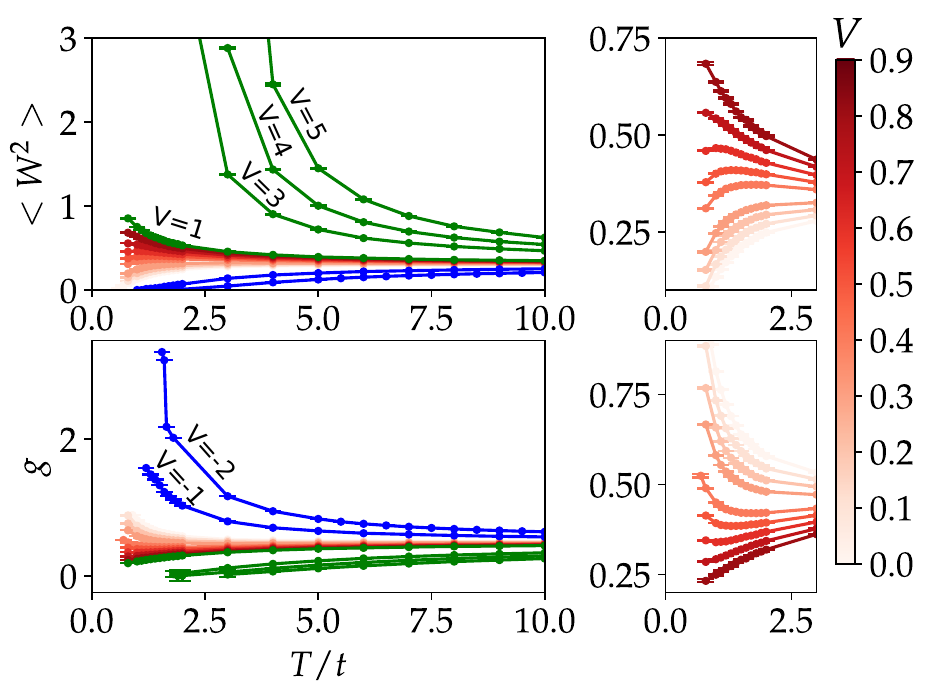}
\caption{\label{fig:w2} Winding fluctuations $\langle W^2 \rangle$ (top) and  Coulomb gas constant $g$ (bottom), as a function of temperature, for different potential values $V/t$. The red curves show values of $0 \le V < 1$ where the reentrance effect is seen, the blue curves show values $V < 0$ and the green curves $V \ge 1$.The corresponding plots to the side show the reentrance region where the coulomb gas constant varies non-monotonically with temperature, one can clearly see this reentrance region is quite small closer to the RK point. This data was obtained from QMC simulations for linear size $L=24$.}
\end{figure}

Fig.~\ref{fig:w2} presents the temperature dependence in the high-temperature phase of winding fluctuations $\langle W^2 \rangle$ (top panel, measured in the QMC simulations) and Coulomb gas constant $g$ (bottom panel, deduced from Eq.~\ref{eq:winding}) for a range of potentials $V/t$. We can distinguish three family of curves. For low potential $V/t < 0$, the winding fluctuations (respectively Coulomb gas constant) monotonously decrease (respectively increases) from their respective $T=\infty$ value ($\langle W^2 \rangle \simeq 0.30343$ and $g=1/2$) when temperature is reduced towards the low-temperature ordered phase (where winding fluctuations vanish).

\begin{figure}[t]
    \includegraphics[width=0.99\columnwidth]{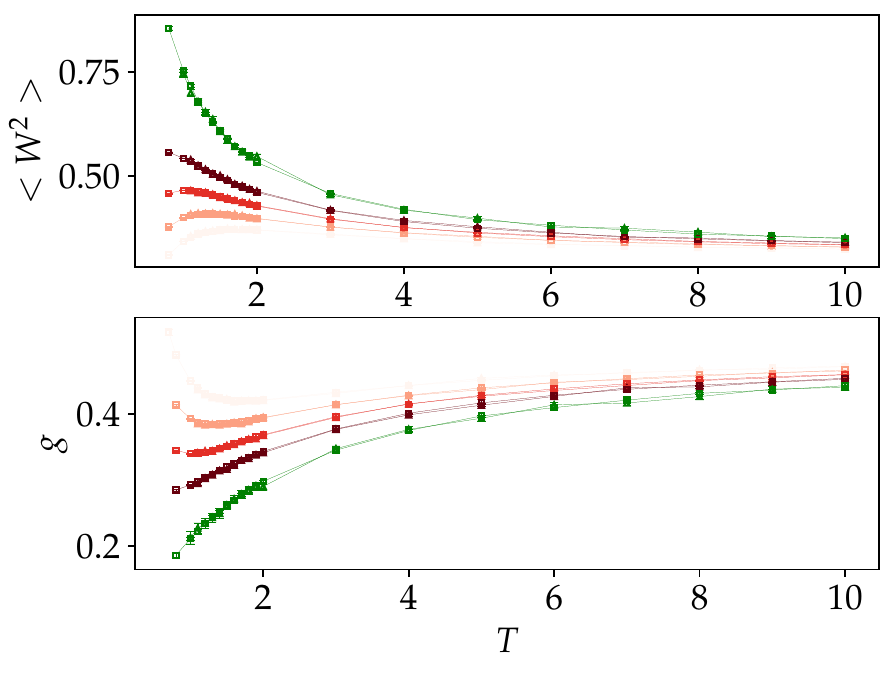}
    \caption{\label{fig:wfinitesize} Variation of winding fluctuations $\langle W^{2} \rangle$ (top plot) and $g$ (bottom plot) with temperature $T$; different colors indicate different values of $V=0.5, 0.6, 0.7, 0.8, 1.0$ (from bottom to top in the top plot). In each case, we show the values for two different sizes $L=24$ and $L=48$.}
\end{figure}

On the other side of the phase diagram, for $V/t > 1$, winding fluctuations (respectively $g$) keep increasing (decreasing) as the temperature is lowered taking us to the staggered phase where the winding numbers are maximal (see Sec.~\ref{sec:results3}). At the RK point ($V/t=1$), we expect the winding fluctuations and $g$ to follow a similar behavior at any finite temperature, even though the $T=0$ limit is singular (as there $g=1/2$). The final family of curves (potential values $0 \lesssim V/t < 1$), show an unexpected re-entrant behavior: winding fluctuations start increasing before eventually showing a downturn at low enough temperature to finally decrease towards their expected behavior in the ground-state (which is long-range ordered forming a columnar or mixed state, for this range of potential). This downturn is most visible for $V/t=0.3, 0.5, 0.6$ in Fig.~\ref{fig:w2} for the system size considered in these simulations. The results for $g$ in Fig.~\ref{fig:w2} with $L=24$ have been interpolated linearly between different values of $V/t$ to produce the lines of constant $g$ in Fig.~\ref{fig:pd}.

We also checked the finite size dependence of $\langle W^{2} \rangle$ and $g$ in Fig.\ref{fig:wfinitesize}. There is hardly any difference between the $L=24$ and $L=48$ data in the temperature range presented in the phase diagram in Fig.~\ref{fig:pd}. Therefore we assume that $L=24$ is sufficient to extract the Coulomb gas constant reliably in this range. This non-monotonous behavior of lines of constant criticality contrasts with the monotonous behavior observed for $g$ in the CIDM and thus disagrees with the finite-temperature phase diagram conjectured in Ref.~\onlinecite{Alet2006}. In Appendix~\ref{sec:app-highT}, we present an argument based on high-temperature expansions of the QDM and CIDM that can account for the shape of the lines of constant criticality at (high enough) temperature. 

Besides the fluctuations of winding, the Coulomb gas constant $g$ also determines (in part) the power-law behavior of correlators. As discussed in Sec.~\ref{sec:obs2}, dimer-dimer correlations contain a vertex contribution  decaying as $x^{-1/g}$, and a dipolar one decaying as $x^{-2}$ in the critical phase. In the re-entrant region where $g<1/2$, the dimer-dimer correlations should exhibit a leading $x^{-2}$ decay, even though $g$ changes with temperature. On the other hand, monomer-monomer correlators do not have a dipolar contribution, and are predicted to decay as $x^{-g}$. An exactly similar phenomenon has been observed in a bipartite classical dimer model where longer-range dimers are included~\cite{Sandvik2006}. 

\begin{figure}[t]
\includegraphics[width=0.99 \columnwidth]{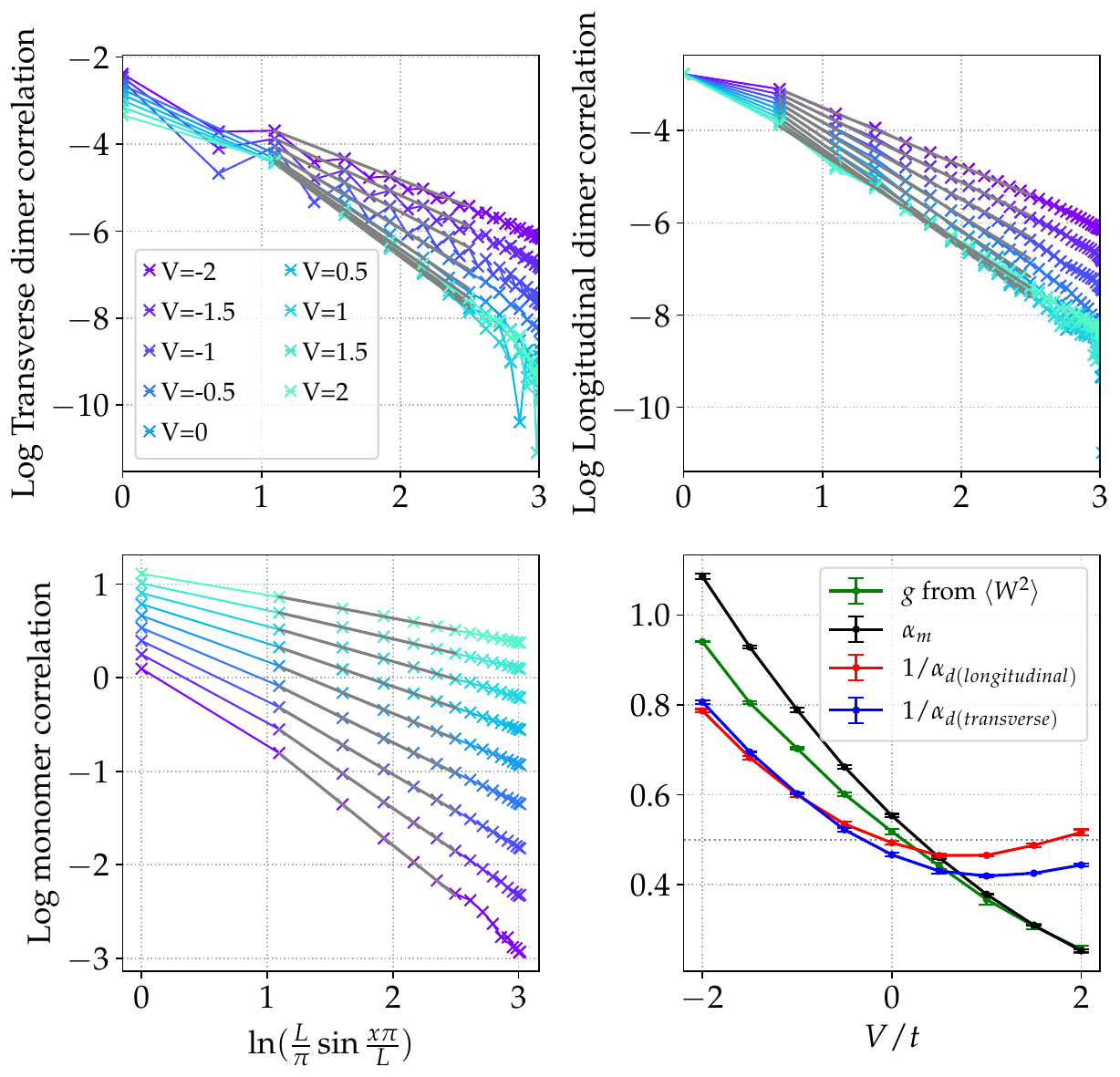}
\caption{\label{fig:correlators} Dimer and monomer correlations at different values of $V/t$ in the high-temperature regime of the quantum dimer model. Panels a,b,c: Transverse dimer, longitudinal dimer and monomer correlators as a function of the conformal distance $L/\pi \sin (\pi x /L)$ plotted in a log-log scale to illustrate their power-law behavior. The grey solid lines represent power-law fits in the decay range. In panel a, one can clearly see the odd-even effect present in the transverse dimer correlations for short distances; we note that the fits were done for even sites. Panel d: Comparison of the Coulomb gas constant $g$ obtained from winding number fluctuations with the exponents obtained from the real space decay of dimer and monomer correlators, as a function of $V/t$. Data are obtained at a fixed temperature $T/t=4$ and for a system size $L=64$.
}
\end{figure}

\begin{figure}[t]

\includegraphics[width=0.99 \columnwidth]{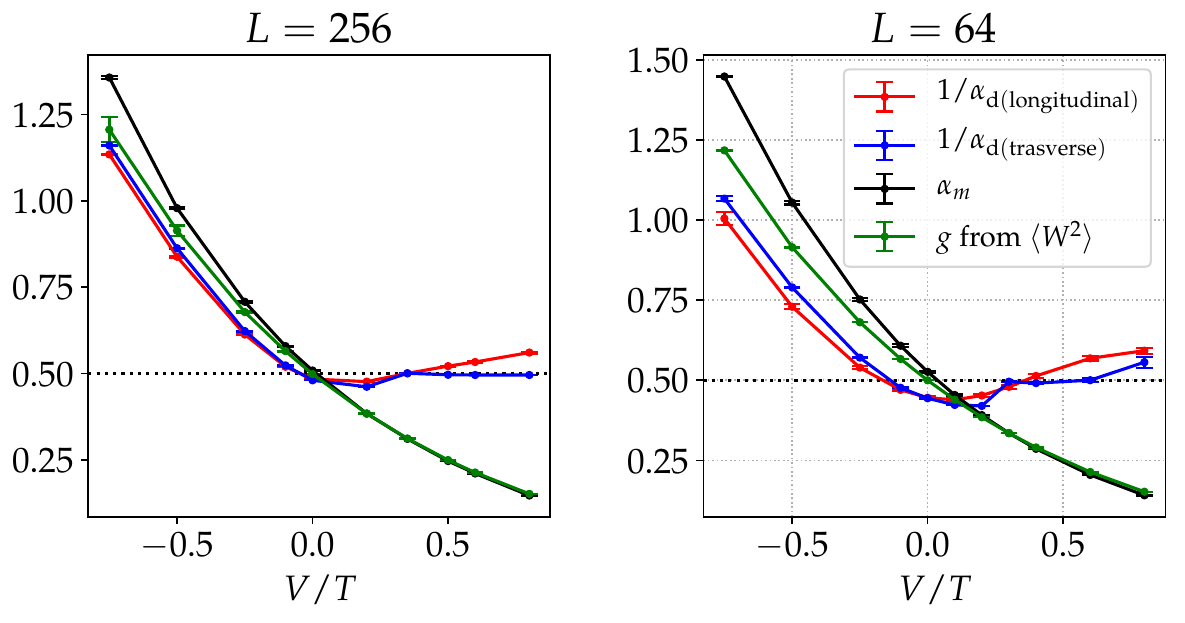}
\caption{\label{fig:correlators_classical} Similar to panel d of Fig.~\ref{fig:correlators}, but for the classical interacting dimer model. Left and right panels correspond to estimates obtained for systems of linear sizes $L=256$ and $L=64$ showing that finite-size effects affect more strongly the estimates of $g$ from real-space decay of dimer and monomer correlators than from winding fluctuations. Simulations have been performed using the classical directed loop algorithm discussed in Ref.~\onlinecite{Alet2006}.}

\end{figure}

We summarize our findings for the power-law decay of correlators in Fig.~\ref{fig:correlators}. Panels a, b and c display the real-space behavior of the transverse dimer-dimer $G^t(x)$, longitudinal dimer-dimer $G^l(x)$ and monomer-monomer $M(x)$ correlators respectively, as a function of different values of $V/t$ for a fixed intermediate to high-temperature $T/t=4$ and a sample of linear size $L=64$. Solid grey lines represent power-law fits in their range of evaluation. The resulting exponents are presented in the last panel $d$, with the deliberate choice of plotting the inverse of the exponent of the dimer-dimer correlators, in order to compare to estimates of $g$ obtained from winding fluctuations and from the exponent of the monomer-monomer correlator decay. We find an overall reasonable agreement between different estimates (from winding fluctuations, decay of $M(x)$ and of $G^t(x),G^l(x)$) of $g$ when $g>1/2$, whereas for $g<1/2$, the inverse of the exponent of dimer-dimer correlator decay tend to flatten close to the expected value $1/2$. We note that the agreement between different estimates is close, but not perfect. We attribute this to finite-size effects, which we demonstrate by reporting the same analysis in Fig.~\ref{fig:correlators_classical} for the CIDM for which we can perform simulations on larger systems. Indeed, we find (Fig.~\ref{fig:correlators_classical}) that the estimates for $L=64$ display some differences, while estimates for $L=256$ are much closer. Comparing data from $L=64$ and $L=256$ in the CIDM shows that the estimate of $g$ obtained from winding fluctuations at $L=64$ is closer to the one at $L=256$ than the ones obtained through correlators. This is the reason why the values of $g$ presented in Fig.~\ref{fig:pd} have been estimated through winding fluctuations of a system of moderate size $L=24$. We also further checked that for a given fixed value of $\langle W^2 \rangle$, real-space decays of correlators in the QDM or CIDM are almost in perfect agreement.

\subsection{The $V/t>1$ range : the transition to the staggered regime }
\label{sec:results3}

\begin{figure}[t]
\includegraphics[width=0.99 \columnwidth]{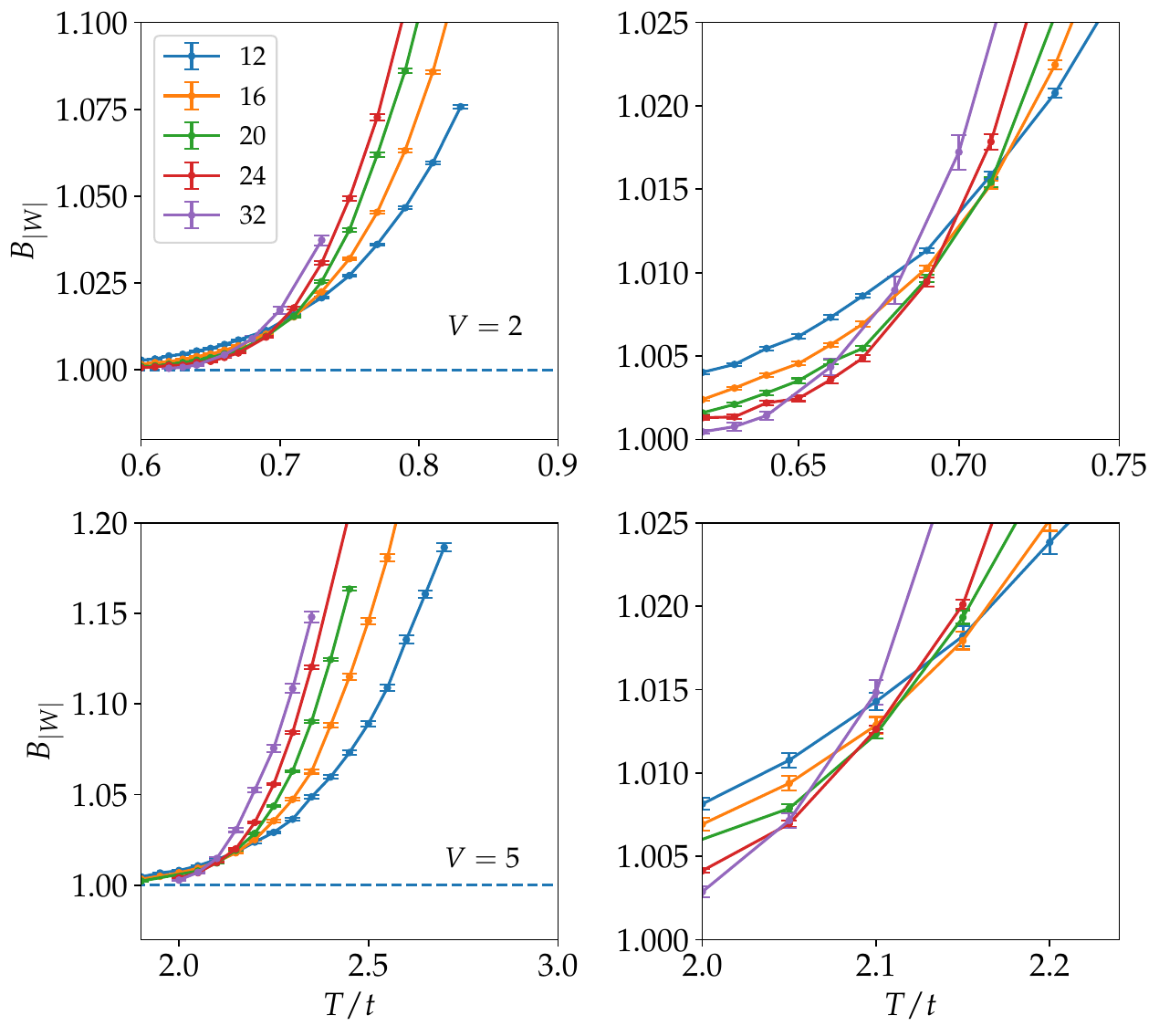}
\caption{\label{fig:Binderstaggered} Binder crossing $B_{|W|}$ at the staggered transition for $V=2$ (top) and $V=5$ (bottom) showing the Binder cumulant $B_{|W|}$ approaches $1.0$ (dashed line) at low temperatures. On the right of each figure we show a zoom near the pseudo-crossing point.}
\end{figure}

We conclude our analysis by showing results for the transition into the staggered phase at low temperature for potential $V/t>1$. We display the temperature dependence of the absolute winding Binder cumulant $B_{|W|}$ for two values of potential in Fig.~\ref{fig:Binderstaggered}. We find a clear decrease of the Binder cumulant from its high-temperature behavior (where $B_{|W|}\rightarrow 4$, not shown) to its low-temperature behavior ($|B_{|W|}\rightarrow 1$) signaling a finite-temperature phase transition to the staggered phase. 

\begin{figure}[h]
\includegraphics[width=0.99 \columnwidth]{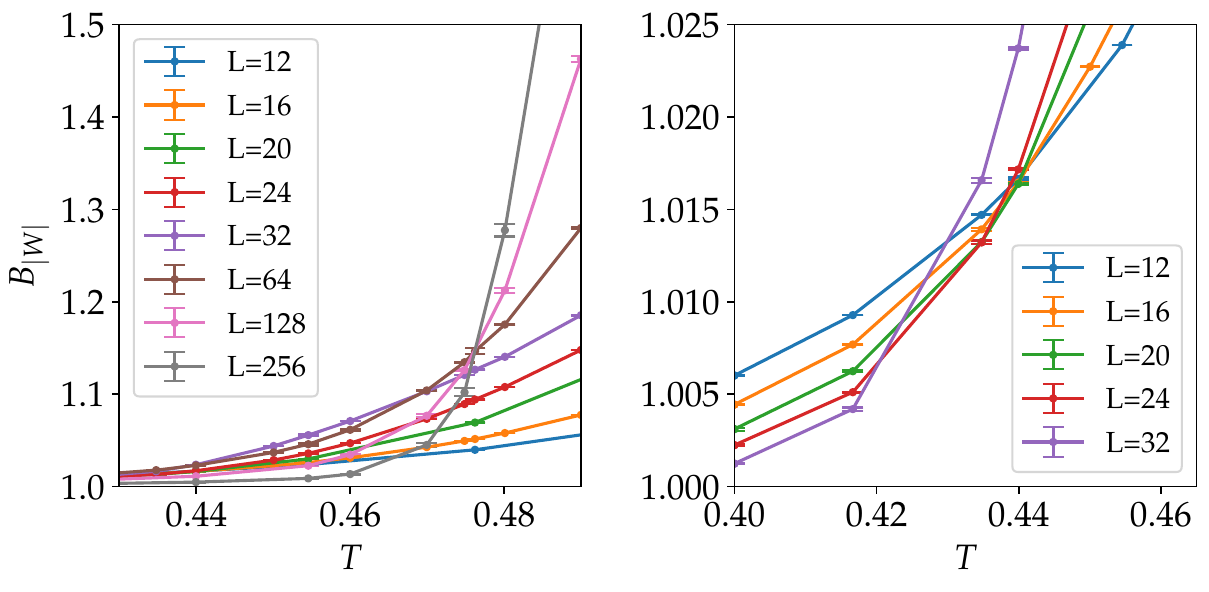}
\caption{\label{fig:Binderstaggered_classical} Binder crossing $B_{|W|}$ at the staggered transition for the classical interacting dimer model. Left panel: for large systems up to $L=256$ (data in agreement with Ref.~\onlinecite{Wilkins2020}). Right panel: zoom on the pseudo-crossing point for smaller systems $L\leq 32$, showing the similarity with the QDM data of Fig.~\ref{fig:Binderstaggered}.}
\end{figure}

The right panels of Fig.~\ref{fig:Binderstaggered} display a zoom in the critical region where an approximate crossing of curves for different system sizes is observed: while data for $L=12,16,20$ appear to cross at a single point, the crossing point is shifted towards slightly lower temperatures for $L=24$ and $L=32$. We argue that this is a finite-size effect that would disappear if one would be able to reach larger system sizes (which are currently out of reach for the QDM, even with our improved QMC algorithm). Indeed, we report in Fig.~\ref{fig:Binderstaggered_classical} the same behavior for the Binder cumulant $B_{|W|}$ for the CIDM (with repulsive interactions $V>0$) on systems of similar size $L=12 \dots 32$, as well as simulations on larger systems (which are possible on the CIDM) which clearly show a better crossing point. We note that the critical temperature is actually slightly larger (by about 10\%) when considering large system sizes than the one obtained from consideration of $L=12..32$ data. We assume that a similar small shift in the staggered transition temperature should also occur for the QDM, which is why the corresponding estimate in Fig.~\ref{fig:pd} is presented with an increased error bar.

The very similar behavior (note the same vertical scales in the main panels of Fig.~\ref{fig:Binderstaggered} and Fig.~\ref{fig:Binderstaggered_classical}) of $B_{|W|}$ for two values of potential in the QDM and the CIDM suggest that the finite-temperature phase transitions to staggered phase are in the same universality class for both models. This phase transition has been argued to be of second-order type in Ref.~\onlinecite{Wilkins2020}.

\section{Conclusion}
\label{sec:conc}

Our main results, presented in the form of the phase diagram of Fig.~\ref{fig:pd}, reveal an overall good agreement between the finite-temperature phase diagram of the QDM and of the classical interacting dimer model~\cite{Alet2006,AletHDR,Wilkins2020} with a fully critical high-temperature phase and transitions into ordered phases of the same type as in the CIDM. This is of course not unexpected. 
What was less expected, is to find lines of Coulomb gas constant $g<1/2$ (typical of repulsive interactions in the CIDM) in the part of the phase diagram with values of $V/t$ at which the ground-state is mixed/columnar ordered (corresponding to flat phases in the height language).
Within the range of parameters (temperature, system size) explored in our simulations, we find that the $g=1/2$ curve follows asymptotically, but never crosses, the $V/t=0$ line. This indicates that the nature (repulsive or attractive) of potential terms dictate the high-temperature behavior even though the potential strength $V$ is much smaller than kinetic terms $t$. This asymptote is well reproduced by the high-temperature expansion argument presented in Appendix~\ref{sec:app-highT}, which also matches fairly well the numerical constant $g$ lines in a fair temperature range.

While the existence of a critical regime up to infinite temperature is not valid for magnetic systems for which the QDM is an effective theory, we nevertheless expect, as argued in Ref.~\cite{Castelnovo2006}, that the extended critical regime found in the QDM can be observed in a large swathe of finite temperature in realistic models where kinematic constraints and frustration play an important role.  

Several interesting perspectives are directly opened up by our work. First, it would be interesting to try to understand analytically the form of the finite-temperature critical lines of critical exponent that we observe numerically, starting from $T=0$ {\it e.g.} through perturbative calculations around the RK point. This could allow to understand whether the observed re-entrant behavior is connected to the U$(1)$-like behavior observed at low energies near the RK point~\cite{Banerjee2014,Oakes18}, as well as to make a connection with the critical lines obtained through high-temperature expansions of Appendix~\ref{sec:app-highT}.

Next, the idea of promoting efficient classical Monte Carlo moves and algorithms, as presented in Appendix~\ref{sec:appendix-MC}, to path integral simulations in complement to the sweeping cluster update should allow the exploration of the finite-temperature phase diagram of a large number of quantum constrained models. For instance, an interesting extension would be to understand the finite-temperature melting of the ``devil's staircase''~\cite{Fradkin2004} of commensurate and incommensurate phases that is argued to be present in a region of the zero-temperature phase diagram of a QDM on the honeycomb lattice~\cite{Schlittler2015}. Also exploring the finite-temperature phase diagrams of the QDM on the cubic or diamond~\cite{Sikora2009,Sikora2011,Bergman2006} lattices, where a Coulomb phase is present, is certainly of great interest.

Finally, our algorithm at the moment does not allow to probe the finite-T phase transition in the range $V/t \in ] 0, 1[$ for the QDM Eq.~\ref{eq:HQDM}, as the critical temperatures are too low. Improving the ability to reach even lower temperatures while keeping Monte Carlo simulations fully ergodic remains an important algorithmic challenge for the future.

\begin{acknowledgments}
We thank Stephen Powell and Gr\'egoire Misguich for insightful discussions, and in particular the latter for kindly providing us the high-temperature expansion argument that form the basis of Appendix B. This work benefited from the support of the joint PhD program between CNRS and IISER Pune, 
and the grant NanoX n$^{\circ}$ ANR-17-EURE-0009 in the framework of the ``Programme des Investissements d’Avenir". We acknowledge the use of HPC resources from CALMIP (grants 2021-P0677 and 2022-P0677) and GENCI (projects A0090500225 and A0110500225). We also acknowledge the use of computing resources in NSM Param Brahma (IISER Pune) during the algorithm development.
\end{acknowledgments}

\pagebreak
\bibliography{QDM}

\appendix
\section{Sweeping cluster and equal time updates}
\label{sec:appendix-MC}
Based on the Stochastic Series Expansion (SSE)~\cite{Sandvik2010} framework for quantum Monte Carlo computations, the Sweeping Cluster Algorithm (SCA) \cite{Yan2019} is an update method for constrained systems such as the quantum dimer model. 
SSE methods~\cite{Sandvik2010} estimate the thermal expectation values of observables by stochastic summation of the series
\begin{equation}
\langle \hat{O} \rangle = \frac{1}{{Z}}{\rm Tr}[{e^{-\beta H}}\hat{O}]=\frac{1}{Z}\sum_{\psi\in \mathcal{D}} \sum_{n=0}^\infty \frac{\beta^n}{n!} \langle \psi |(-H)^n\hat{O}| \psi \rangle \label{eq:expandTrExp}
\end{equation}
where, in the present context, $\psi$ is summed over all fully packed hard core dimer configurations $\mathcal{D}$. 
The simulated Hamiltonian $H=H_{QDM}-c$ is defined by 
\begin{equation}
-H = \sum_{i \in \mathcal{P} } (\hat{K}_i + \hat{V}_i)
\end{equation}
where $\mathcal{P}$ is the set of all the $N_{\mathcal{P}}$ plaquettes on the lattice, and $c$ is a constant (a parameter of the algorithm) such that $c+v\geq 0$ (we use the notation $v = -V$ for this appendix only).  The kinetic term ``vertex'' $K_i$ on a plaquette has non-zero matrix elements between two dimer configurations which differ only in the $i$-th plaquette; the non-zero matrix elements being 
\begin{equation}
\langle\parallelH|K_i|\parallelV\rangle = \langle\parallelV|K_i|\parallelH\rangle = 1.\nonumber
\end{equation}
The potential term ``vertex'' $V_i$ is diagonal in the basis of classical dimer configurations and its diagonal matrix elements depend on $i$-th plaquette in the following way 
\begin{equation}
\langle\parallelV|V_i|\parallelV\rangle = \langle\parallelH|V_i|\parallelH\rangle = c+v\nonumber
\end{equation}
and
\begin{multline}
\langle\emptyPlaq|V_i|\emptyPlaq\rangle = \langle\PlaqOIH|V_i|\PlaqOIH\rangle = \langle\PlaqIOH|V_i|\PlaqIOH\rangle =\\= \langle\PlaqOIV|V_i|\PlaqOIV\rangle = \langle\PlaqIOV|V_i|\PlaqIOV\rangle = c.
\end{multline}
Assuming that the summation Eq.~\ref{eq:expandTrExp} over $n$ can be truncated at some finite value $M$, and expanding the powers of the Hamiltonian in terms of $K_i$ and $V_i$, we can rewrite thermal expectation value as follows:
\begin{equation}
\langle \hat{O} \rangle=\frac{1}{Z}\sum_{\psi\in \mathcal{D}} \sum_{w\in \mathcal{W}} \frac{\beta^{l_w}}{(M-l_w)!} \langle \psi |w| \psi \rangle O_\psi
\label{expectationO}
\end{equation}

Here $\mathcal{W}$ is the set of all ordered operator sequences of lengths less than or equal to $M$ that can be made from the elements of the set $\{K_i | i\in \mathcal{P}\}\cup \{V_i | i\in \mathcal{P}\}\cup \{\mathbb{I}\}$; and $\mathbb{I}$ is the identity operator.
The symbol $l_w$ is the number of non-identity elements (i.e. not $\mathbb{I}$) in the sequence $w$.
For simplicity, in this discussion, we assume that the observable ${O}$ is diagonal in the dimer basis with an eigenvalue $O_\psi$ in the dimer configuration $\psi$. SSE samples the pairs $(\psi,w)$ from the following probability distribution over $\mathcal{D}\times \mathcal{W}$ 
\begin{equation}
{P}(\psi,w)\propto\frac{\beta^{l_w}}{(M-l_w)!}\langle \psi |w| \psi \rangle.\label{eq:dist}\nonumber
\end{equation}
Thermal expectation value of $\hat{O}$ can now be estimated as the expectation value of $O_\psi$ in ${P}$.

The cutoff length $M$ in Eq \ref{expectationO} 
is chosen such that it is larger than the maximum number of non-identity operators $l_{w}$ that appeared at any iterations in the past, making sure that the truncation in the expansion causes no errors. In practice, the following procedure is routinely executed within the Stochastic Series Expansion method ~\cite{Sandvik2010}: during the equilibration steps, if it is found that $l_w> f M$  in any of the iterations (where $f$ is an arbitrary parameter, chosen as $f=0.8$ in our simulations), $M$ is incremented by $M+\Delta M$ (such that $M>l_w/f$) by inserting identity vertices at random positions in the string.

The pairs $(\psi,w)$ with non-zero probability $P$ can be visualized as a ``state evolution'' in which the initial state $\phi_0=\psi$ evolves through the states $\phi_t \sim (\prod_{i=1}^t w_i)\phi_0$  at ``time slice'' $t$, ultimately back to the initial state $\phi_{t=M}=\psi$. This identification means that there are periodic boundary conditions in imaginary time, see Fig. \ref{fig:spatial}~(a,c) for illustration. A time slice $t$ is then associated with state $\phi_t$ and an operator $w^{(t)}$.

Dimer constraints severely restrict the allowed configurations rendering most simple strategies to sample $(\psi,w)$ very inefficient. The Sweeping cluster algorithm~\cite{Yan2019} presents an efficient Monte Carlo method, based on the SSE framework, to sample within the constrained space. We first quickly summarize the SCA here for completeness. This is followed by a discussion of the strategy to improve the ergodicity of the algorithm by performing an additional equal time update.

\subsection*{Sweeping Cluster Algorithm}

As usual in SSE methods~\cite{Sandvik2010}, the SCA has two components: a standard diagonal update that only modify time slices with  identity $\mathbb{I}$s or diagonal $V$ vertices, and an off-diagonal update (the sweeping cluster update) which results in interchanges between $K_i$s and $V_i$s vertices, as well as some transitions in $\psi$. 

Starting from a valid $(\psi,w)$, this sweeping cluster update changes some of the $K_i$'s in the sequence $w$ into $V_i$'s and vice-verse while making the requisite changes in $\psi$ such that $(\psi,w)$ is still valid. 
The reader is referred to Ref.~\onlinecite{Yan2019} to have a detailed explanation of the exact procedure.

\subsection*{Equal-time directed loop update}
\label{sec:equal-time}

\begin{figure*}
\includegraphics[width=1.0\textwidth]{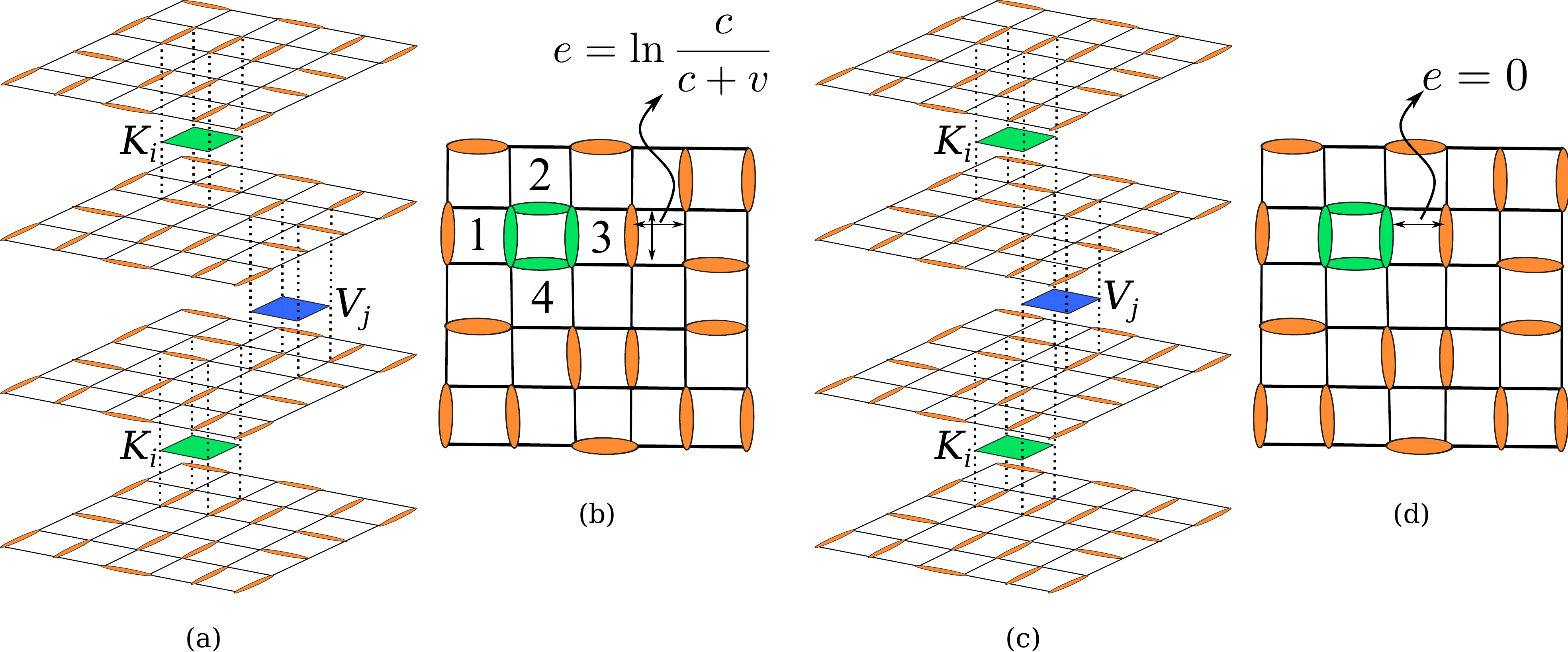}
\caption{Illustration of the construction of the equal-time update. The green and blue squares in the panels (a) and (c) represent $K_i$ and $V_j$ vertices respectively. Panel (a) illustrates an operator string in which the diagonal $V_j$ vertex is located in a plaquette not adjacent to the plaquette containing the $K_i$ vertex. In the effective model, there is a pseudo-energy $e=\ln\frac{c}{c+v}$ associated to parallel dimers inside the plaquette $j$. Dimers in other plaquettes do not interact. The effective model is shown in panel (b). The edges around plaquette containing $K_i$ are fully filled and frozen. In the plaquettes nearest to it, labelled 1,2,3,4 in panel (b), the edges touching the plaquette containing $K_i$ are always empty. Panel (c) illustrates an operator string which has plaquettes containing $V_j$ and $K_i$ adjacent to each other. The rules giving the effective model is similar to that in panels (a,b) except that the blue squares ($V_j$) contribute to the pseudo-energy $e$ in $j$ only if the edge shared by the two plaquettes is occupied in the time slice containing $V_j$. In the illustration the edge is unoccupied, so $e=0$.
\label{fig:spatial}}
\end{figure*}

The sweeping cluster update  is found to be not ergodic in practice; this is indicated by the absence of winding number fluctuations in the samples generated from the sweeping cluster algorithm. We rectify this using the equal time update which uses a directed loop update in $\psi$, similar to the one used for classical dimer or link models~\cite{Alet2006,Sandvik2006,Alet2003}. 
It is easy to see that any plaquette $i$ that contains a non-diagonal vertex ($K$) in any of the time slices has to be flippable in $\psi$. A generic directed loop algorithm however will not preserve flippability of the plaquettes. To circumvent this problem, we can perform directed loop updates in a restricted region within the state $\psi$. The construction of this equal-time update is presented below and illustrated in Fig.~\ref{fig:spatial}.

Given the operator sequence $w$, we identify the set of all plaquettes $\mathcal{P}_K$ which have a non-diagonal vertex $K$ in any time slice (we will represent the set of other plaquettes by ${\overline{\mathcal{P}}}_K$).  The equal time update allows transitions within a set of states which are identical in the plaquettes in $\mathcal{P}_K$, 
while satisfying detailed balance (corresponding to the  distribution $P$ in Eq. \ref{eq:dist})

To do this, we create a new state ${\psi'}$ which is identical to $\psi$ but with dimers occupying all edges around any plaquette in $\mathcal{P}_K$ (represented by green dimers in Fig. \ref{fig:spatial}(b,d)). We can now associate infinite fugacity for dimers on these edges such that these edges are unaffected by the directed loop updates on ${\psi}'$ (that is, the dimer configuration will not be modified on these edges).

This however is insufficient to guarantee detailed balance. On any plaquette in $\overline{P}_K$ that contains a diagonal vertex $V$ (in any time slice), the weight associated with a flippable plaquette in $\psi$ is $c+v$ whereas it is $c$ for a non-flippable plaquette. To satisfy detailed balance, the directed loop should take into account the locations of the $V$ vertices. We do this by associating a pseudo-energy ${E}$ for configuration ${\psi}'$ with an interaction between parallel dimers in plaquettes in  $\overline{\mathcal{P}}_K$ as described below.

\begin{enumerate}
\item If there are no diagonal vertices ($V$) in any time slice of a given plaquette, there is no pseudo-energy ($e_i=0$) associated to this plaquette $i$.
\item Now we consider any plaquette $i$ which is {\it not adjacent} to any plaquette in $\mathcal{P}_K$.
If there are $n$ diagonal vertices ($V$s), at various time slices on $i$, we associate a pseudo-energy of ${e}_i=n\ln(\frac{c}{c+v})$ between any two parallel dimers in $i$. This is illustrated in Fig \ref{fig:spatial}(a,b).

\item Now consider the case where there are diagonal vertices $V$ on a plaquette $i$ which is {\it adjacent} to one of the plaquettes $k$ in $\mathcal{P}_K$. There is one edge shared by $k$ and $i$. Notice that the occupancy of this edge is different in different time slices. Let $n_2$ be the number of diagonal vertices $V_i$ that occur in time slices where the {\it shared edge is occupied}. The pseudo-energy associated with parallel dimers in $i$ is ${e}_i = n_2 \ln(\frac{c}{c+v})$. Only one of the four edges in this plaquette can change its state during the directed loop update (the one which is opposed to the shared edge). This is illustrated in Fig \ref{fig:spatial}(c,d).
\item If there is a plaquette $i$ adjacent to more than one plaquettes from $\mathcal{P}_K$, then none of the edges in $i$ can change their occupancy, and thus it is not necessary to define its pseudo-energy.
\end{enumerate}
Thus we obtain a pseudo-energy ${E}=\sum_{i\in \mathcal{P}} {e}_i x_i$ where $x_i=1$ if there is a parallel pair of dimers in the $i-$th plaquette, $x_i=0$ otherwise. Note that this is a classical interacting dimer model similar to the ones studied in Ref. \onlinecite{Alet2005,Alet2006}, and we can thus use the directed loop algorithm for classical dimers from Refs.~\onlinecite{Alet2006,Sandvik2006,Alet2003}. Performing this directed loop update starting from ${\psi}'$, new configurations ${\psi}'_1$ which satisfy detailed balance under $\bar{P}\propto e^{-{E}}$ are produced. Finally the new state $\psi_1$ in the pair $(\psi_1,w)$ is constructed by copying the configurations of $\mathcal{P}_K$ in $\psi$ into ${\psi}'_1$.

The directed loop updates are efficient at sampling dimer configurations $\psi$; winding number sectors are effectively sampled if $\mathcal{P}_K$ is not large. In practice, we can reach equilibration of winding numbers up to intermediate temperatures on fairly large systems, see the main manuscript for typical range of temperature and sizes.\\

%

\section{High temperature expansion}
\label{sec:app-highT}

\begin{figure*}[h]

\includegraphics[width=0.99 \textwidth]{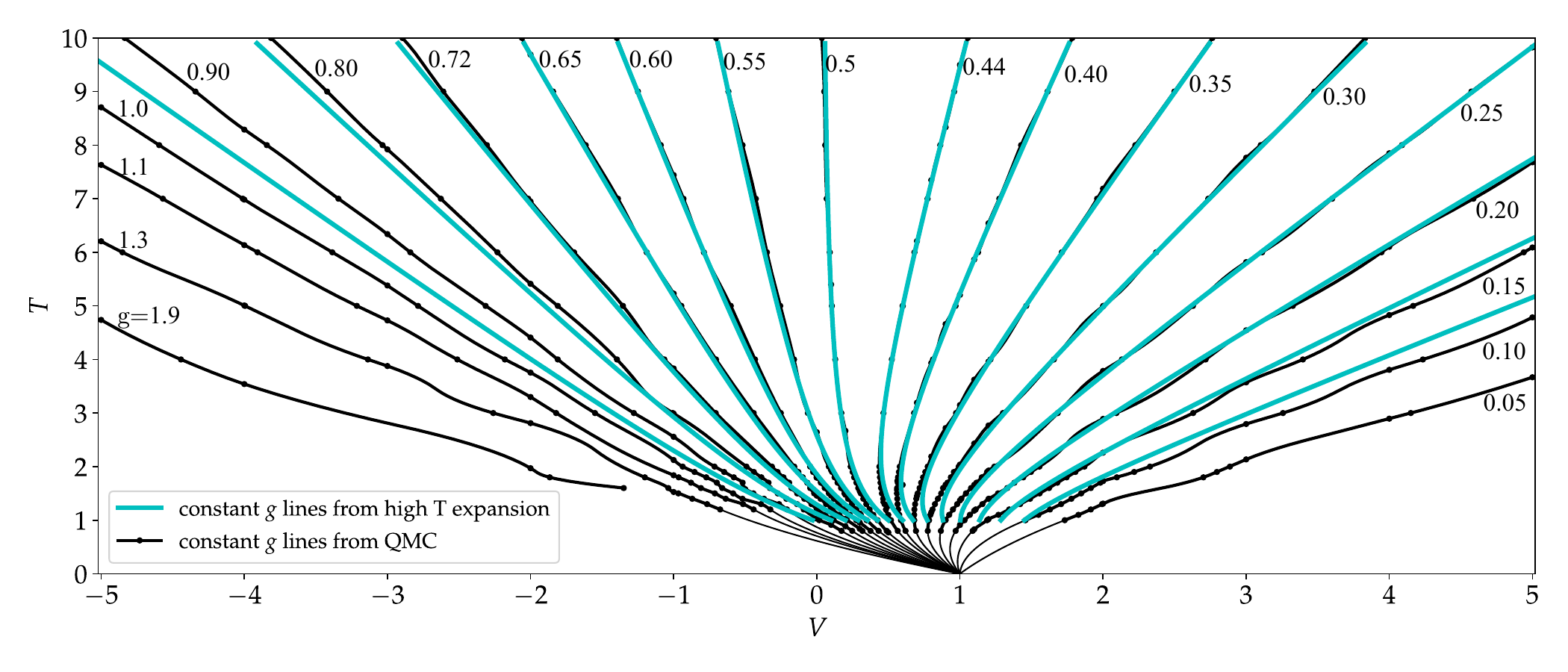}

\caption{\label{fig:pd2} Comparison between the critical constant-$g$ lines obtained from QMC (black lines, same legend as displayed in Fig.~\ref{fig:pd}), and those  predicted from the high-temperature expansion of Appendix~\ref{sec:app-highT} (cyan).}
\end{figure*}

We present here an argument and a computation that can explain the behavior of the constant-$g$ lines in the high-temperature regime. We do this by comparing a high-temperature expansion for the expectation values of observables in the QDM and in a CIDM, allowing us to identify the parameters of an effective CIDM that mimics the QDM. We are very grateful to G. Misguich for providing us this argument.  

We consider the QDM Hamiltonian Eq.~\ref{eq:HQDM} (with parameter $V$ and at temperature $T$) as well as the (effective) CIDM with classical energy $E_c=W N_c^f$ which associates an energy $W$ to any plaquette with two parallel dimers in the dimer configuration $c$. Here and below $N^f_c=N^c(\begin{tikzpicture}[scale=0.35,baseline={([yshift=-0.6ex]current bounding box)}]
    \draw (0,0) rectangle (1,1);
    \draw[fill=Dandelion] (0.5,1) ellipse (0.56 and 0.12);
    \draw[fill=Dandelion] (0.5,0) ellipse (0.56 and 0.12);
\end{tikzpicture})+N_c( 
\begin{tikzpicture}[scale=0.35,baseline={([yshift=-.6ex]current bounding box)}]
    \draw (0,0) rectangle (1,1);
    \draw[fill=Dandelion] (0,0.5) ellipse (0.13 and 0.56);
    \draw[fill=Dandelion] (1,0.5) ellipse (0.13 and 0.56);
\end{tikzpicture}
)$ is the total number of such flippable plaquettes in $c$. We use the symbol $W$ for this energy scale to distinguish it from $V$ of the QDM.

Let us consider an operator ${\cal O}$ which is diagonal in the dimer configuration basis, that is $\langle c | {\cal O} | c \rangle={\cal O}_c$ and expand its (unnormalized) expectation value in the QDM in powers of inverse temperature $\beta=1/T$: 
\begin{multline}
{\rm Tr} [{\cal O} e^{-\beta H_{\rm QDM}}] \simeq  {\rm Tr}[{\cal O}]  - \beta {\rm Tr} [{\cal O} H_{\rm QDM}]+\dots\\+\frac{\beta^2}{2} {\rm Tr} [{\cal O} H_{\rm QDM}^2] + O(\beta^3)
\end{multline}
%
Since the observable is diagonal in the computational basis, it is easy to see that ${\rm Tr}[{\cal O}H_{\rm QDM}]=\sum_c {\cal O}_c V N^f_c$ and ${\rm Tr} [{\cal O}H^2_{\rm QDM}]=\sum_c {\cal O}_c \left[ V^2 (N^f_c)^2+ t^2 N^f_c \right] $. We finally obtain:
\begin{multline}
{\rm Tr} [{\cal O} e^{-\beta H_{\rm QDM}}] \simeq  \sum_c O_c  +\dots \\ +\sum_c O_c N^f_c \left[ -\beta V + \beta^2 \frac{t^2}{2} + \beta^2 \frac{V^2}{2} N^f_c \right]+ O(\beta^3)
\end{multline}.

The high temperature expansion for the CIDM gives
\begin{multline}
\sum_c {\cal O} e^{-\beta E_c} \simeq  \sum_c O_c  +\dots \\ + \sum_c O_c N^f_c \left[ -\beta W+ \beta^2 \frac{W^2}{2}  N^f_c \right]+ O(\beta^3).
\end{multline}
Up to an error $O(\beta^3)$, the two expressions match if we identify $W$ with $V-\beta t^2/2$. This means that the physics (including $e.g.$ the value of $g$) at a high-temperature point with coordinates $(V,T)$ in the phase diagram Fig.~\ref{fig:pd} for the QDM (setting $t=1$) is same as that of a CIDM with parameters $(W=V-\frac{1}{2T},T)$. In other words
\begin{equation}
g_{\rm QDM}(V,T)\simeq g_{\rm CIDM}(W=V-\frac{1}{2T},T).\label{eq:equivalgqdm-cidm}
\end{equation}
This is what we use to obtain the cyan colored lines in Fig.~\ref{fig:pd2}. We obtain $g_{\rm CIDM}(W,T)$ by simulating the CIDM for a range of parameters in a system of same size $L=24$ (same as the size of the QDM). We then use Eq. \ref{eq:equivalgqdm-cidm} to infer $g_{\rm QDM}$; cyan lines in Fig.~\ref{fig:pd2} represent the constant $g_{\rm QDM} $ contours. 

These contours show an excellent match with the numerical data from direct simulations of the QDM down to temperatures $T\sim 1.5$. 
We note that the negative sign in the second term of the effective value $W=V-1/(2T)$ explains the asymptotic line located to the right of $V=0$ at large temperature for the $g=1/2$ curve (which is predicted to follow a $T=1/(2V)$ behavior) as well as the positive (negative) slopes of the $g<1/2$ ($g>1/2$) lines at high-enough temperature. Indeed results on the CIDM~\cite{Alet2005,Alet2006,Wilkins2020} indicate that repulsive $W>0$ (attractive $W<0$) interactions always lead to a decrease (increase) of $g$ from its infinite-temperature value $g=1/2$. Overall this argument provides a good account of the high-temperature behavior and hints at the re-entrant behavior observed at finite temperature in the $V>0$ region.

\end{document}